\documentclass[12pt]{article}

    \usepackage[T1]{fontenc}
    \usepackage{graphicx,epsfig,amsmath,amsfonts,cite}
    \renewcommand{\abstract}{}
\begin{document}
\title{Optical vibrations  in alkali halide crystals}

\author{Anton Stupka
      }

 \maketitle

\begin{center} {\small\it Oles Honchar Dnipropetrovs'k National University, Gagarin
ave., 72, 49010 Dnipropetrovs'k, Ukraine\\antonstupka@mail.ru }
\end{center}

\begin{abstract}  We consider long-wave  phonon-polaritons and longitudinal optical
phonons in alkali-halide ionic crystals. The model of point
charges that are polarized in the self-consistent electromagnetic
field in a dielectric environment is used. The standard dispersion
laws for both branches of phonon-polaritons and longitudinal
optical phonons are obtained. The transversal optical phonon
frequency is found from the electrostatic equilibrium condition.
It is proved by comparison with tabular data that the found
frequency coincides with the ion plasma frequency multiplied on
the relation
$\sqrt{(\varepsilon_{\infty}+2)/(3(\varepsilon_0-\varepsilon_\infty))},$
where $\varepsilon_{\infty}$ and $\varepsilon_0$ are the
high-frequency dielectric constant and the static one
respectively.

 keywords{ Alkali-halide ionic crystal;  self-consistent
electromagnetic field; long-wave  oscillations; phonon-polariton;
longitudinal optical phonon; electrostatic equilibrium; dielectric
constant.}

PACS{Phonons in crystal lattices, 63.20. e, Electrostatic waves
and oscillations plasma waves, 52.35.Fp,  Collective effects
71.45.-d}
\end{abstract}

\maketitle


\section{Introduction}

The simplest and best understood of the crystals containing two
different kinds of atoms are unquestionably the alkali halides.
The positive and negative ions lie on two interpenetrating cubic
lattices. According to the Born theory of ionic crystals (1),
there are ions with unit (positive or negative) charge, the
binding being largely due to the Coulomb interaction between the
ions, and the crystal being stabilized by short-range repulsive
forces between nearest neighbors. The importance of taking into
account an electromagnetic interaction when considering long-wave
optical vibrations in ionic solids is shown in (2). Deep analysis
of early theories was proposed in (3). Often an effective charges
is introduced to satisfy the experimental data for optical
frequencies. The
 Szigeti effective charge is essentially less than
elementary one for the $NaCl$-type alkali halides  (4). In recent
works ab intio  calculations are performed to find the dynamical
matrix which allows to calculate the phonon frequencies of an
ionic crystal (5),(6).

The long-wave acoustical properties of an alkali halide crystal
are closely related with electrostatic interaction in it and
transversal acoustic phonon velocity in the such crystal is
defined by the electrostatic energy (7).   We will show that
optical phonon frequencies are defined via electrostatic
properties of the alkali-halide crystal too. In the recent paper
(8) the consideration of the high-frequency optical vibrations in
ionic crystals with two atoms per unit cell as plasma oscillations
of point charges  is proposed. Further presentation puts designed
to transfer the results of (8) to low-frequency limit of
phonon-polaritons.

\section{Transversal optical phonon frequency }

Lets consider  long-wave compared with a lattice constant
high-frequency oscillations in the diatomic ionic crystals  that
allows using the macroscopic consideration. We will consider ions
as point charges but use a high-frequency permittivity
   $
\varepsilon _ \infty $ to describes the electronic (atomic)
polarization. Then each ion becomes a dipole in the external
electric field.
 As well know,
 when an ion of each node is the center of cubic symmetry that
 allows to use the  Lorentz relation for a local field.
 The local field acting on the ions is expressed through the mean field
 as follows (27.30) from (9):
 \begin{equation}\label{lf}
    \mathbf{E}^{loc}=(\varepsilon_{\infty}+2)/3 \mathbf{E}.
\end{equation}

Elastic force is proportional to the displacement gradients that
in long-wave approximation is neglected. Thermal motion of ions a
fortiori is neglected, since the velocity of thermal motion less
then velocity of acoustic waves. Damping will not be taken into
account. Shall consider small fluctuations in non-magnetic media,
then one can omit the nonlinear magnetic part of the Lorentz
force. So we write the linearized equations of motion for ions in
this model

\begin{equation}\label{1}
\partial {\bf v}_ +  /\partial t = -\omega_0^2({\bf u}_ +-{\bf u}_ -)+e{\bf E}^{loc}/M_ +,
\end{equation}
\begin{equation}\label{2}
\partial {\bf v}_ -  /\partial t = -\omega_0^2({\bf u}_ --{\bf u}_ +)  - e{\bf E}^{loc}/M_ -
.
\end{equation}
This sign $ \pm $
  corresponds to a charge, $
M_\pm $
   is a mass, respectively, of positively and negatively charged
   ions,
   $\omega_0$ is a resonant frequency (27.47) from (9).
   The full  derivative  with respect to time   coincides with
the partial one after linearization. It is convenient to introduce
a relative displacement of sublattices ${\bf u}={\bf u}_+-{\bf
u}_-.$ As easy to see, we can obtain from equations [\ref{1}] and
[\ref{2}]
\begin{equation}\label{12}
\partial^2 {\bf u}  /\partial t^2 =-\omega_0^2{\bf u}+ e(\varepsilon_{\infty}+2)/(3M){\bf
E},
\end{equation}
where  a reduced mass of a crystal unit cell  $ M = {{M_ + M_ -}}/
{({M_ + + M_ -})} $ is introduced.

Then we can write an ionic (displacement) polarization density in
the linearized case
\begin{equation}\label{P}
 {\bf P}_i=en_0 {\bf u},
\end{equation}
where ions are considered as point charges with
  the equilibrium density $ n_0 .$    Now
we obtain the relation for the dielectric induction for an
isotropic case that is linearly related to the  electric field
strength $ {\bf E} $ in the approximation of small oscillations
(10),(11).
\begin{equation}\label{5}
{\bf D}  = \varepsilon_\infty  {\bf E}+4\pi  {\bf P}_i.
\end{equation}
On the other hand,  we have the relation for the dielectric
induction via the dielectric constant $\varepsilon_0 $ for the
static situation
\begin{equation}\label{50}
{\bf D}_0  = \varepsilon_0  {\bf E}.
\end{equation}
Then we obtain a relation between field ${\bf E}$ and displacement
${\bf u}$ from the equations [\ref{5}] and [\ref{50}] in the
static case
\begin{equation}\label{5stat}
\varepsilon_0  {\bf E}=\varepsilon_\infty  {\bf E}+4\pi en_0 {\bf
u}.
\end{equation}
The second relation between field ${\bf E}$ and displacement ${\bf
u}$  can be obtained from the equation [\ref{12}] in the static
case
\begin{equation}\label{12stat}
0 =-\omega_0^2{\bf u}+ e(\varepsilon_{\infty}+2)/(3M){\bf E}.
\end{equation}
That two equations allow us to find the resonant frequency
$\omega_0$ without finding the force matrix only from the
condition of electrostatic equilibrium:
\begin{equation}\label{fr2}
\omega_0  =\sqrt{\frac{4\pi
e^2n_0(\varepsilon_{\infty}+2)}{3M(\varepsilon_0-\varepsilon_\infty)}}
.
\end{equation}

As well known, the coefficient $\omega_0$ in the [\ref{12}] is
correspond to the transversal optical phonon frequency (36.12)
from (12).

To compare with tabulated values of frequency $ \omega _0^{tab} $
 it is  convenient  in the expression [\ref {fr2}] to pass from the density of ions of the same
 sign to the mass density of the crystal
   $
\rho $: $ \frac {{n_0}} {M} = \frac {\rho} {{M_ + M_ -}} $. Then
we can write

\begin{equation}\label{24}
\omega _0  = {\rm 1}{\rm .70156} \cdot \sqrt {\frac{\rho
(\varepsilon_{\infty}+2)}{{3  M_ +  M_ -
(\varepsilon_0-\varepsilon_\infty)}}} \,10^{ - 9} \,c^{ - 1}.
\end{equation}

For comparison, we use the data of Table 2.2. from (13) for
longitudinal oscillation frequencies $ \omega _0^{tab}.$ Values of
density $ \rho $ of ionic crystals are taken from (15).

\begin{table}[h!]
\noindent \caption{ Transversal optical frequencies of some alkali halide crystals }%
\vskip3mm\tabcolsep4.2pt

\noindent{\footnotesize
\begin{tabular}{c c c c c c c c}
 \hline%
 \multicolumn{1}{c}{\rule{0pt}{9pt}Crystal}%
 & { $ \rho
\,$}g/cm $^{  3}$ & { $ \varepsilon_0$} &{ $ \varepsilon_\infty$}
 & { $
\omega _0^{tab} 10^{13}  $}Hz
 & { $
\omega _0 \,10^{13} $}Hz & { $ \omega _0/\omega _0^{tab} $}
\\%
\hline%

\rule{0pt}{9pt}  LiF& 2.64 &8.9& 1.9 &5.783&6.25  & 1.08 \\NaF&
 2.79&5.1&1.7
&4.502& 4.93&1.10
\\NaCl& 2.17&5.9&2.25& 3.089&3.29&1.07 \\NaBr&    3.21&
6.4&2.6& 2.524 &2.72&1.08\\KCl& 1.99&4.85&2.1&
  2.675
&2.74&1.02\\KBr  &2.75&4.9&2.3&2.129  & 2.26&1.06\\KI
&3.12&5.1&2.7&1.902 & 2.08&1.09
\\RbCl &2.76&4.9&2.2&2.185& 2.23&1.02 \\RbBr &2.78&4.9&2.3&1.658& 1.54&0.93 \\RbI &3.55&5.5&2.6&1.413& 1.35&0.95
\\CsCl&    3.97&7.2&2.6  & 1.865&
1.72&0.92\\CsBr &4.44&6.5&2.8&1.375& 1.35&0.98 \\
\hline

\end{tabular}
}\label{tabl}
\end{table}

As shown in the Table~\ref {tabl}, we have a good coincidence of
values obtained from the formula [\ref {24}] with known ones (13).
The Szigeti  expression (14) (see, also, (2.67) from (13))
contains an extra factor $\sqrt{(\varepsilon_{\infty}+2)/3}$
comparing with [\ref{fr2}] that worsens the agreement with
experimental data.

\section{ Phonon-polaritons and longitudinal optical phonons}

   The self-consistent mean electromagnetic field must satisfy Maxwell's equations in a dielectric.
\begin{equation}\label{3}
\partial {\bf D}/\partial t = crot{\bf B},
\end{equation}
\begin{equation}\label{4} \partial {\bf B}/\partial t =  -
crot{\bf E}.
\end{equation}

Then we have a  homogeneous  system of time equations [\ref {12}],
 [\ref {3}] and [\ref {4}] and the coupling equations
[\ref {lf}], [\ref{P}] and [\ref{5}] for high-frequency long-wave
vibrations of the ionic lattice and the self-consistent
electromagnetic field.

Then we can obtain an equation for electric field waves from
mentioned equations system. Lets take the derivative with respect
to time of the equation [\ref {3}] and substitute the derivatives
$\partial {\bf B}/\partial t$ from  [\ref{4}]
\begin{equation}\label{14} \partial^2 \varepsilon _\infty  {\bf
E}/\partial t^2 = -c^2\nabla\times(\nabla\times{\bf E}) - 4\pi
en_0
\partial^2 {\bf u}/\partial t^2.
\end{equation}

 In obtained equations  it is convenient to pass
to the Fourier-components by the following rule
\begin{equation}\label{19} {\bf E}\left( {{\bf x},t} \right) =
\smallint d^3 kd\omega {\bf E}\left( {{\bf k},\omega }
\right)e^{i{\bf kx} - i\omega t} /(2\pi )^4 .
\end{equation}
Then  we obtain from
 the [\ref {12}]
 \begin{equation}\label{12fu}
{\bf u}  =
\frac{e(\varepsilon_{\infty}+2)}{(\omega_0^2-\omega^2)3M}{\bf E}.
\end{equation}

 Lets divide the field into potential and vortical parts  $ {\bf E}
= {\bf E} ^\parallel  + {\bf E} ^\bot  $. Then  we obtain  linear
homogeneous algebraic equations
\begin{equation}\label{21}  - \omega ^2  {\bf
E}^\parallel   =  - \frac{4\pi  e^2
n_0(\varepsilon_{\infty}+2)}{3\varepsilon _\infty M}{\bf
E}^\parallel \frac{\omega ^2 }{\omega ^2 -\omega_{0} ^2 }.
\end{equation}

\begin{equation}\label{20}  -
\omega ^2 {\bf E}^ \bot   =  - \frac{c^2}{\varepsilon _\infty} k^2
{\bf E}^ \bot    - \frac{4\pi  e^2
n_0(\varepsilon_{\infty}+2)}{3\varepsilon _\infty M} {\bf E}^ \bot
\frac{\omega ^2 }{\omega ^2 -\omega_{0} ^2 },
\end{equation}

  From [\ref {21}] omitting trivial $\omega=0$ it is obtained a longitudinal oscillations frequency
\begin{equation}\label{23} \omega _L^2  =\omega_{0} ^2+ 4\pi  e^2
n_0(\varepsilon_{\infty}+2)/(3 M\varepsilon _\infty),
\end{equation}
that matches to the longitudinal phonons. We can rewrite
[\ref{23}] using [\ref{fr2}]
\begin{equation}\label{232} \omega _L^2  =\frac{4\pi
e^2n_0(\varepsilon_{\infty}+2)}{3M}\left({\frac{1}{(\varepsilon_0-\varepsilon_\infty)}+
\frac{1}{\varepsilon_\infty}}\right)=\omega_{0}
^2\frac{\varepsilon_0}{\varepsilon_\infty},
\end{equation}
that is the famous Lyddane-Sachs-Teller formula (16).

From the equation [\ref {20}] we obtain the upper phonon-polariton
dispersion law for the high-frequency $\omega \gg\omega_{0}$ case
\begin{equation}\label{22} \omega ^2  = c^2 k^2
/\varepsilon _\infty   + 4\pi  e^2 n_0 (\varepsilon_{\infty}+2)/(3
M\varepsilon _\infty).
\end{equation}
That result is analogous to obtained in (8). Solution [\ref{22}]
 became the photon branch for large $ k $. On the other hand,
the relation [\ref{22}] is the dispersion law for transversal
plasma waves (10) in ionic plasma with a multiplier
$\sqrt{(\varepsilon_{\infty}+2)/3
 }$.

In the general case we have from [\ref{20}]
\begin{equation}\label{222} \omega ^4 -\omega ^2(\omega_{L} ^2+c^2
k^2 /\varepsilon _\infty) + \omega_{0} ^2 c^2 k^2 /\varepsilon
_\infty   =0.
\end{equation}
that coincides with (12.6) from (17) and gives both branches of
phonon-polaritons. Now we can see that [\ref{222}] has the
solutions (12.7) from (17)

\begin{eqnarray}\label{sol}
 \omega^2=\frac{1}{2}( \omega _L^2+{c^2}{} k^2/\varepsilon
 _\infty\pm\\
   \pm\sqrt{(\omega _L^2+{c^2} k^2/\varepsilon _\infty)^2-4\omega_{0}
^2{c^2} k^2/\varepsilon _\infty})\nonumber
\end{eqnarray}
The lower branch for short waves gives the expected limit
$\omega\rightarrow \omega_0$ for transversal optical phonon
frequency [\ref{fr2}].

\section{Conclusion}
Then, the transversal optical phonon frequency in an ionic crystal
is found without using any empirical fits. Of course, an ionicity
of compounds is not absolute and  measurement accuracy of used
dielectric constants is not very high. According to the Table~\ref
{tabl}, then we have a good coincidence of values obtained from
the formula [\ref {24}] with known ones.  Correct dispersion laws
for both branches of phonon-polaritons and longitudinal optical
phonons are obtained. Given consideration generalized the work
(8), where ions regarded as free charges for high frequencies.


\begin{thebibliography}{99}
\bibitem{(1)}(1) M. Born, K. Huang, { Dynamical theory of crystal
lattices} (Clarendon, Oxford, 1958).

\bibitem{(2)}(2) Kun Huang, {Proc. Roy. Soc.}
{\bf A208}, 352 (1951).








\bibitem{(3)}(3)
A. D. B. Woons, W. Cochran, B. N. Bnoczaousa,  { Phys. Rev.} {\bf
119}, N 3, 980 (1960).
\bibitem{(4)}(4)
B. G. Dick, JR,  A. W. Overhauser,  {Phys. Rev.}  {\bf 11}, N 1,
90 (1958).

\bibitem{(5)}(5) Priya Sony, Alok Shukla,
 {Phys. Rev. B}  {\bf 77}, 075130 (2008).
\bibitem{(6)}(6)
Yi Wang, Shunli Shang, Zi-Kui Liu,  Long-Qing Che,  {Phys. Rev. B}
{\bf  85}, 224303 (2012).
\bibitem{(7)}(7) A.A. Stupka,  {Ukr. J. Phys.}  {\bf 58},
N 12, 1156  (2013).

\bibitem{(8)}(8)  A.A. Stupka,  {Ukr. J. Phys.}  {\bf 58},   N 9, 863 (2013).





\bibitem{(9)}(9)  N.W. Ashcroft,
N.D. Mermin,   {Solid state physics} (Cengage Learning, Inc , New
York, 1976).
\bibitem{(10)}(10)  Electrodynamics of Plasma, edited by A.I. Akhiezer, (Nauka, Moscow, 1974, in Russian).
\bibitem{(11)}(11) L.D. Landau, E.M. Lifshitz, L.P. Pitaevskii,
 {Electrodynamics of Continuous Media}. Vol. 8 (2nd ed.),
(Butterworth-Heinemann, 1984).


\bibitem{(12)}(12)  O. Madelung,  {Theory of Solid} (Nauka, Moscow, 1980, in Russian).


\bibitem{(13)}(13) J.A. Reissland,  {The physics of phonons} (John Wiley
and sons LTD, London-New York-Sydney-Toronto, 1973).
\bibitem{(14)}(14) B. Szigeti, Trans. Far. Soc., {\bf 45}, 155 (1949).
\bibitem{(15)}(15)  V.A. Rabinovich, Z.Ya. Havin,  {A brief chemical
directory} (Himiya, Leningrad, 1978, in Russian).
\bibitem{(16)}(16) R. H. Lyddane, R. G. Sachs, and E. Teller,  {Phys. Rev.} {\bf 59}, 673
(1941).
\bibitem{(17)}(17) A.S. Davydov,  {Theory of Solids} (Nauka,
Moscow, 1976, in Russian).

\end{thebibliography}
\end{document}